# Normalization of direct citations in publication-level networks: Evaluation of six approaches

Peter Sjögårde[a,b,x], Per Ahlgren [c,y]

[a]Health Informatics Centre, Department of Learning, Informatics, Management and Ethics, Karolinska Institutet, Stockholm, Sweden
[b]University library, Karolinska Institutet, Stockholm, Sweden
[c]Department of Statistics, Uppsala University, Uppsala, Sweden

ORCID:
[x]https://orcid.org/0000-0003-4442-1360
[y]https://orcid.org/0000-0003-0229-3073

Email: peter.sjogarde@ki.se; per.ahlgren@uu.se

Corresponding author: Peter Sjögårde, University Library, Karolinska Institutet, 17177 Stockholm, Sweden

## Abstract

Clustering of publication networks is an efficient way to obtain classifications of large collections of research publications. Such classifications can be used to, e.g., detect research topics, normalize citation relations, or explore the publication output of a unit. Citation networks can be created using a variety of approaches. Best practices to obtain classifications using clustering have been investigated, in particular the performance of different publication-publication relatedness measures. However, evaluation of different approaches to normalization of citation relations have not been explored to the same extent. In this paper, we evaluate five approaches to normalization of direct citation relations with respect to clustering solution quality in four data sets. A sixth approach is evaluated using no normalization. To assess the quality of clustering solutions, we use three measures. (1) We compare the clustering solution to the reference lists of a set of publications using the Adjusted Rand Index. (2) Using the Sihouette width measure, we quantity to which extent the publications have relations to other clusters than the one they have been assigned to. (3) We propose a measure that captures publications that have probably been inaccurately assigned. The results clearly show that normalization is preferred over unnormalized direct citation relations. Furthermore, the results indicate that the fractional normalization approach, which can be considered the standard approach, causes inaccurate assignments. The geometric normalization approach has a similar performance as the fractional approach regarding Adjusted Rand Index and Silhouette width but leads to fewer inaccurate assignments. We therefore believe that the geometric approach may be preferred over the fractional approach.

## 1 Introduction

Constructing classifications of research publications by the use of clustering in citation networks is an efficient way to detect research topics or, at a more aggregate level, research specialties in very large publication collections. Such classifications provide possibilities to study the research landscape. Bibliometric studies of citation networks have a rather long history, starting over half a century ago (e.g., de Solla Price, 1965; Garfield et al., 1964). Large publication-level classifications have been around for about 10 years. Several papers have investigated best practices for clustering of publications (e.g., Ahlgren et al., 2020; Ahlgren & Jarneving, 2008; Boyack & Klavans, 2010, 2020; Klavans & Boyack, 2017; Sjögårde & Ahlgren, 2018, 2020; Velden et al., 2017; Waltman et al., 2020). However, to our best knowledge, no evaluation of normalization approaches to direct citations has been performed. Evaluation of such approaches is the focus of this paper.

In 2012, Waltman and van Eck proposed a methodology to construct an hierarchical publication-level classification of research publications in a large citation network (Waltman & van Eck, 2012). The development of modularity-based optimization algorithms and improved computational capacity had made such approaches possible (Newman, 2004; Newman & Girvan, 2004). The initial modularity-based approaches have been improved during the last decade, both regarding efficiency and the quality of the obtained clustering solutions (Traag et al., 2011, 2019; Waltman & van Eck, 2013).

In the field of scientometrics, quite a lot of research has been devoted to comparing clustering solutions obtained using different publication-publication relatedness measures. Direct citations have been compared to indirect approaches that use co-citations and bibliographic coupling (Boyack et al., 2020; Boyack & Klavans, 2020; Klavans & Boyack, 2017; Waltman et al., 2020, 2020). Expanding the direct citation approach using citations external to a publication set of interest has been shown to increase the quality of clustering solutions (Ahlgren et al., 2020; Boyack & Klavans, 2014). There are also indications that global models perform better than local models (Boyack, 2017). However, none of these studies investigate the use of different approaches to normalization of raw citation-based measures (like number of co-citations).

Normalization of citation relations has mostly been discussed in the context of journal citation networks and indirect citation relations. It was early recognized that the size of journals influences the number of relations (Narin et al., 1972). Leydesdorff (1987) clustered journals using Pearson's $r$ to normalization of co-citation relations. The use of Pearson's $r$ was, though, criticized by Ahlgren et al. (2003), who pointed out drawbacks of this approach. Boyack et al. (2005) compared different relatedness measures in a journal citation network using the Web of Science subject categories as a baseline. With respect to inter-citation frequencies, they preferred the Jaccard normalization approach based on its scalability, the resemblance of the resulting clusters with the Web of Science subject categories and an assessment of visualizations created by the use of different normalization approaches. However, they underscore that the cosine approach to normalization performed just as well as the Jaccard normalization in statistical terms.

In publication-level networks, normalization of direct citations has not been much discussed, and to our best knowledge, no study has (as indicated above) evaluated the use of different normalization approaches to direct citations in large-scale networks of this kind. Waltman and van Eck (2012) proposed a normalization approach that normalize each citation relation with the total number of relations of the publication (see Section 4.1.2). This approach has been used in several studies (Ahlgren et al., 2020; Boyack & Klavans, 2020; Sjögårde, 2022; Sjögårde & Ahlgren, 2018, 2020). However, it has been recognized that clustering methodologies sometimes create loosely connected clusters and results that are less intuitive or even undesirable (Held, 2022; Held et al., 2021; Held & Velden, 2022).

In this contribution, we restrict the analysis to direct citations. Other approaches such as extended direct citations, bibliographic coupling or textual similarity would probably be preferable in a real analysis setting of the datasets used by us, because of the modest sizes of the datasets and the high number of publications with few relations (cf. Section 3 below). However, the sole purpose of our analysis is to compare the performance of different approaches to normalization of direct citations, and we are particularly interested in how the approaches perform in sparse areas of the networks. We evaluate six approaches to normalization of direct citations with respect to clustering solution quality.

The paper is organized as follows. In the next section, we explain the purpose of normalizing direct citation relations when clustering publications. Furthermore, we describe the most commonly used approach, and introduce an observed problem related to this approach. In Section 3, we present the data used in the study. The methods section (Section 4), treats the investigated normalization approaches, the clustering approach used in this study, and the evaluation methodology. We present the results of the study in Section 5. In the last section, we discuss the results and present some conclusions.

## 2 Direct citations: the need for normalization, and a normalization problem

Clustering of publications in citation networks is influenced by some properties of citations. First of all, a citation is a directed relation between two publications. A citation occurs when one (newer) publication refers to an (older) publication. Older publications generally have more citations than newer publications, since they have had more time to be cited. Consequently, older publications generally have more citation relations in a direct citation network. Secondly, the distribution of citations over publications is highly skewed, meaning that a low share of the publications receive a high share of the citations, and a large share of the publications receive few or no citations. Lastly, the referencing practices vary between fields, regarding both the average quantity of references in the publications and the age of the referenced literature. These variations result in different density of citation relations among fields. If no normalization is performed, it is likely that clustering procedures are biased towards old publications, highly cited publications, and fields with high density of citation relations (Sjögårde, 2023).

Normalization has been performed to correct for the biases indicated above. Waltman and van Eck (2012) proposed a normalization procedure that normalizes the citation relation between two publications (say $i$ and $j$) to the total number of relations of $i$. We refer to this approach as "Fractional" approach (for further description, see Section 4.1.2). The fractional approach is probably the most widely used approach for normalization of direct citation relations within the field of scientometrics. Nonetheless, to the best of our knowledge, it has not been empirically assessed and compared to other approaches. Furthermore, we have reasons to believe that this approach may not be the best alternative, at least in some circumstances. In the following, we will describe such a circumstance and illustrate a problem, which is related to the fractional normalization approach and which we address in this paper.

Figure 1 shows a cluster belonging to a clustering solution obtained by Sjögårde (2022). The fractional approach to normalization of direct citations and the Leiden clustering algorithm, the latter in combination with the Constant Potts Model quality function, were used to obtain the solution in a large direct citation network of publications. Nodes represent publications, and node size corresponds to total number of citation relations (including relations outside the cluster). Edges represent citation relations, and edge thickness corresponds to edge weights, here based on normalized citation relations.

Consider the node in red color in the top of the figure, say $i$. This node has 51 relations in total, but only one of these within the cluster. The relation within the cluster has a very high weight, however, because the node related to the red one, say $j$, has only a few relations, namely four of which three falls within the cluster. In the fractional approach, the edge weight for $i$ and $j$ is equal to $(1/4+1/51)/2 \approx 0.13$, i.e., the average of the two normalized citation relations[1]. Further, the other 50 nodes related to $i$ have 23 to 3,437 relations. Let us denote the node with 23 relations as $k$. The edge weight for $i$ and $k$ is approximately 0.03 $((1/23+1/51)/2)$. This means that the edge weight for $i$ and $j$ is about four times higher than the weight between $i$ and $k$. In the clustering process, the high relation weight for $i$ and $j$ yields that the clustering algorithm is rewarded (with respect to the quality function) for assigning $i$ and $j$ to the same cluster. We find it concerning, though, that cluster membership of publications with many relations can be determined by one or a few publications with a small number of relations. It can be noted that 12 of the publications related to *i* belong to another cluster. This cluster may be a better alternative for the assignment of *i*.

---

[1] Alternatively, the denominator can be dropped.

Now, one way to reduce relative differences of the indicated kind would be to normalize a direct citation against the geometric mean of the total number of relations of the two publications. With regard to the nodes *i*, *j* and *k*, this yields an edge weight equal to $1/\sqrt{4 \times 51} \approx 0.07$ for *i* and *j*, whereas the corresponding weight for *i* and *k* is equal to $1/\sqrt{23 \times 51} \approx 0.03$. By using a geometric mean approach, the edge weight for *i* and *j* is about 2.3 times higher than the corresponding weight for *i* and *k*, a substantial reduction from four. Indeed, the geometric mean approach is one of the approaches to normalization of direct citations that we evaluate in this work.

We need to point out that the cluster in Figure 1 should not be seen as representative, but as an illustration of a problem that seem to occur in some of the small clusters. However, the extent of this problem is hard to estimate since it is not easily measured.

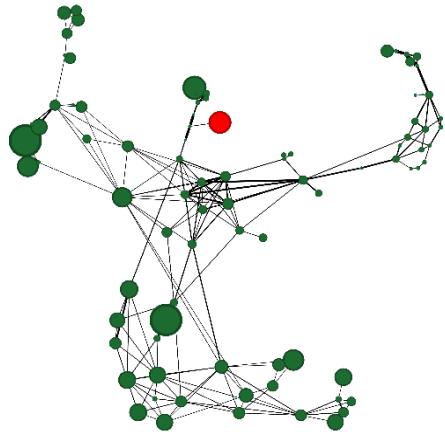

*Figure 1: The fractional approach. An example of a cluster with a node (in red color) with many relations outside the cluster and exactly one relation within the cluster.*

## 3 Data

We used four sets of publications for the evaluation of approaches to normalization of direct citations between publications. Each set was retrieved by searching the in-house version of PubMed/MEDLINE at Karolinska Institutet for a Medical Subject Heading (MeSH). The MeSH terms were selected from different branches in the MeSH tree, and we aimed to pick terms with high semantic dissimilarity. Publications were retrieved for each MeSH term, including their sub-terms. If the MeSH term was located in several places in the tree, we used sub-terms from all branches. We only considered terms as "major topics" (terms marked with asterisks in the PubMed web interface). The following MeSH terms were used: "Psychology, Social" (we write "Social Psychology" in the remainder of this paper), "Autoimmune Diseases", "Metabolism" and "Stem Cells". We restricted each search to the publication years 1995-2021. Each of the four terms retrieves a rather large set of publications, ranging from about 160,000-440,000.

The NIH Open Citation Collection (iCite et al., 2019) was used to retrieve citation relations between the publications in each set. We considered the relations as undirected, and we removed duplicates: if publication *i* cites publication *j* and *j* cites *i*, we only took one of these relations into account. Table 1 shows descriptive statistics for the four datasets.

*Table 1: Descriptive statistics for the four datasets.*

| Dataset | # Publications | # Publications with at least 1 relation | # Citation relations | Avg. relations/ Publication |
|---|---|---|---|---|
| Social Psychology | 386,204 | 271,879 | 1,595,274 | 4.1 |
| Autoimmune Diseases | 298,157 | 269,486 | 3,692,263 | 12.4 |
| Metabolism | 437,717 | 397,328 | 2,955,214 | 6.8 |
| Stem Cells | 162,093 | 151,950 | 2,527,699 | 15.6 |

The distribution of relation counts over publications is highly skewed in all four datasets (Table 2). Most publications have few relations and a small proportion of the publications have more than 100 relations. Only a few publications have more than 1,000 relations. Figure 2 shows the distribution of relations in the four datasets as box plots with violin wrapping. In the Social Psychology set, there is a dense concentration of publications with only 1-5 relations. The datasets Autoimmune Diseases and Stem Cells are not as highly skewed as the two other data sets, and the publications in these sets generally have more citation relations.

*Table 2: Number of publications with 1-10, 11-100, 101-1000 and >1000 relations respectively.*

| Dataset | 1-10 relations | 11-100 relations | 101-1000 relations | > 1000 relations |
|---|---|---|---|---|
| Social Psychology | 174,392 | 94,999 | 1,486 | 2 |
| Autoimmune Diseases | 93,933 | 166,242 | 9,254 | 57 |
| Metabolism | 226,869 | 166,447 | 4,000 | 12 |
| Stem Cells | 42,013 | 102,547 | 7,320 | 70 |

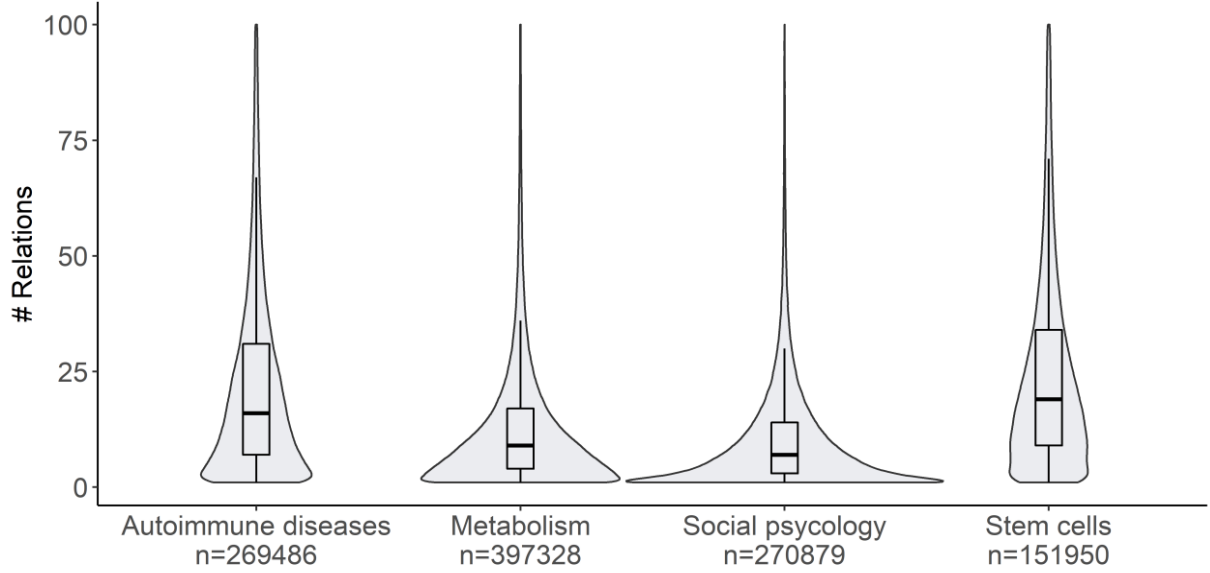

*Figure 2: Box plots with violin wrapping showing the distribution of relations over publications. Restricted to publications with 1-100 relations.*

## 4 Methods

In this section, we first describe the six approaches to normalization of direct citations. We then briefly describe the clustering of the publications, and finally we present our evaluation methodology.

### 4.1. Investigated approaches

The six normalization approaches used in this study give rise to corresponding publication-publication relatedness measures. In the following six subsections, we describe the approaches through the definitions of their corresponding relatedness measures. The seventh subsection puts forward edge weight examples.

*4.1.1. Unnormalized*

The unnormalized relatedness of two publications, *i* and *j*, is defined as (Ahlgren et al., 2020):

$$r_{ij} = \max(c_{ij}, c_{ji}) \qquad (1)$$

where $c_{ij}$ is 1 if *i* cites *j*, 0 otherwise. Thus, if a citation relation exists from either *i* to *j* or from *j* to *i*, then $r_{ij}$ is 1, otherwise 0. Note that $r_{ij}$ is undirected.

*4.1.2. Fractional*

For the fractional approach, we used the definition provided by Waltman and van Eck (2012). The normalized relatedness of *i* with *j* is defined as:

$$a_{ij} = \frac{r_{ij}}{\sum_k c_{ik}} \qquad (2)$$

where $\sum_k r_{ik}$ is the total number of relations of $i$. However, since the network is undirected, we also considered the normalized relatedness of $j$ with $i$ to calculate the edge weight. We use the average of $a_{ij}$ and $a_{ji}$ for the edge weight between $i$ and $j$, i.e. as the normalized relatedness of $i$ and $j$, which we denote by $r_{ij}^{frac}$. $r_{ij}^{frac}$ ranges from 0 to 1.

*4.1.3. Geometric mean*

The geometric mean approach is similar to the fractional approach. However, in this approach we divide $r_{ij}$ with the geometric mean of the total number of relations of $i$ and $j$. The normalized relatedness of $i$ and $j$ is defined as

$$r_{ij}^{geom} = \frac{r_{ij}}{\sqrt{\sum_k r_{ik} \times \sum_k r_{jk}}} \quad (3)$$

$r_{ij}^{geom}$ ranges from 0 to 1.

*4.1.4. Geometric mean-limitN*

This approach is similar to the geometric mean approach but uses a restriction of the minimum value of $\sum_k r_{ik}$ and $\sum_k r_{jk}$ in the calculation. Geometric mean-limitN reduces the edge weight of relations for publications with less than $N$ relations. The normalized relatedness of $i$ with $j$ is defined as:

$$r_{ij}^{geom-limN} = \frac{r_{ij}}{\sqrt{d_{ik} \times d_{jk}}} \quad (4)$$

where $d_{ik} = \sum_k r_{ik}$ if $\sum_k r_{ik} \geq N$, otherwise $d_{ik} = N$. We used $N = 5$, which yields that $r_{ij}^{geom-limit5}$ ranges from 0 to 0.2.

*4.1.5. Directional-fractional*

The directional-fractional approach, as well as the directional-geometric approach defined below, differs from the other approaches in that the direction of the citation relation is taken into consideration when calculating the edge weight (Yun et al., 2020). The normalized relatedness of $i$ and $j$ is defined as

$$r_{ij}^{prob-frac} = \begin{cases} 0 \text{ if } r_{ij} = 0 \\ \left(\frac{r_{ij}}{i_{ref}} + \frac{r_{ij}}{j_{cit}}\right)/2 \text{ if } r_{ij} = 1 \text{ and } i \text{ cites } j \\ \left(\frac{r_{ij}}{i_{cit}} + \frac{r_{ij}}{j_{ref}}\right)/2 \text{ if } r_{ij} = 1 \text{ and } j \text{ cites } i \end{cases} \quad (5)$$

where $i_{ref}$ is the number of references in $i$ and $j_{cit}$ is the number of citations to $j$. The probability that a citation exists from $i$ to $j$ increases with increasing number of references in $i$ and increasing number of citations to $j$. However, the probability of a citation from $i$ to $j$ is not affected by increasing number of references in $j$ or citations to $i$. Therefore, the number of references in $j$ and the number of citations to $i$ is disregarded in the calculation of $r_{ij}^{prob-frac}$. $r_{ij}^{prob-frac}$ ranges from 0 to 1. Even though direction is considered in the definition of the measure, the citation relation is used as undirected as for the other considered measures.

*4.1.6. Directional-geometric*

The directional-geometric approach is basically the same as bidirectional normalization used by Yun et al. (2020). The difference between directional-fractional and directional-geometric is analogous to the difference between the fractional and the geometric approach. Here, the normalized relatedness of *i* and *j* is defined as

$$r_{ij}^{prob-geom} = \begin{cases} 0 \text{ if } r_{ij} = 0 \\ \frac{r_{ij}}{\sqrt{i_{ref} \times j_{cit}}} \text{ if } r_{ij} = 1 \text{ and } i \text{ cites } j \\ \frac{r_{ij}}{\sqrt{i_{cit} \times j_{ref}}} \text{ if } r_{ij} = 1 \text{ and } j \text{ cites } i \end{cases} \quad (6)$$

$r_{ij}^{prob-geom}$ ranges from 0 to 1

*4.1.7. Edge weights across four of the approaches–examples*

Table 3 illustrates, for some example values of the total number of relations of *i* and *j*, how the edge weight varies across the four approaches that do not take direction into consideration.[2] Note that the geometric mean approach, but not the geometric mean-limit5 approach, results in the same weight as the fractional approach if $\sum_k r_{ik} = \sum_k r_{jk}$, i.e. when *i* and *j* have the same number of relations. However, the edge weight is lower using $r_{ij}^{geom}$ compared to $r_{ij}^{frac}$ when the total number of relations differs between *i* and *j*. The variation of edge weights is much smaller for geometric mean-limit5 approach than for the geometric mean approach and for the fractional approach.

*Table 3: Examples of the variation of edge weight using the different normalization approaches. Only approaches that do not take direction into consideration are covered.*

| $\sum_k r_{ik}$ | $\sum_k r_{jk}$ | Unnormalized | Fractional | Geometric mean | Geometric mean-limit5 |
|---|---|---|---|---|---|
| 1 | 1 | 1.00 | 1.00 | 1.00 | 0.20 |
| 10 | 10 | 1.00 | 0.10 | 0.10 | 0.10 |
| 100 | 100 | 1.00 | 0.01 | 0.01 | 0.01 |
| 1 | 10 | 1.00 | 0.55 | 0.32 | 0.14 |
| 1 | 100 | 1.00 | 0.51 | 0.10 | 0.04 |
| 10 | 100 | 1.00 | 0.06 | 0.03 | 0.03 |

## 4.2. Clustering

For each approach and dataset, we obtained a series of clustering solutions using the Leiden algorithm (Traag et al., 2019). The Leiden algorithm was used to maximize the Constant Potts Model as quality function (Traag et al., 2011; Waltman & van Eck, 2012). This model is resolution limit free, which means that it can be used to detect clusters at granular levels. We used the following values of the resolution parameter γ to obtain the clustering solutions: 0.05, 0.01, 0.005, 0.001, 0.0005, 0.0001.

---

[2] To keep Table 3 as simple and illustrative as possible, we have left out the directional approaches.

## 4.3. Evaluation methodology

We compared the six normalization approaches described in Section 4.1 with respect to clustering solution quality. For this, the following three measures were used to evaluate the clustering solutions: (1) The Adjusted Rand Index (ARI), (2) Silhouette width, and (3) Number of probably inaccurate assignments. The evaluation measures are described in Section 4.3.1, whereas result visualization is treated in Section 4.3.2.

### 4.3.1. Evaluation measures

In this section, we describe the three evaluation measures. We doubt that there exists a ground truth for clustering solutions. Our intention is, though, to capture different aspects of clustering solution quality.

### ARI

ARI is a measure of the similarity between two classifications of the same set of objects (Hubert & Arabie, 1985). The measure takes values on the interval [0, l]. We used ARI to compare a baseline solution to clustering solutions obtained using the different normalization approaches. The baseline was created in a similar manner as in Boyack and Klavans (2017) and Sjögårde and Ahlgren (2018). In each of the four datasets of publications, we retrieved those publications that had more than 100 references in total and a minimum of 50% of the references within the dataset. The restriction to 50% is more inclusive than in Sjögårde and Ahlgren (2018), because we used a much smaller dataset. Furthermore, we restricted the set to publications published from 2019 or later. We regard the retrieved set of publications as baseline classes and their references as the items of these classes. The baseline classes were used as proxies for research topics. The classes should not be seen as perfect representations of topics. However, we assume that a large proportion of the references in each baseline publication is likely to be connected to the same topic. We therefore believe that baseline classes can be used as proxies for topics and for comparative purposes.

We followed the same procedure as in Sjögårde and Ahlgren (2018) in order to fulfill requirements of the properties of the baseline classes and the clustering solutions. To avoid having more than one baseline class addressing the same topic, we used the following procedure. Bibliographic coupling was used to calculate the similarity between baseline classes. If the items of two classes had an overlap of 30% or more of their references, we regarded the two classes as addressing the same topic. Based on these same-topic relations, we created groups of connected classes. From each group, we selected one class at random and used this class as a baseline class and excluded the other classes in the same group. A second requirement of the properties of the baseline classes is that the classes must be pairwise disjoint. To fulfill this requirement, we referred each item of the baseline classes to one class only. Each item was referred to the class to which it had the highest frequency of citation relations.

As a final step, we delimited each clustering solution (obtained using a dataset and a normalization approach) to the publications represented in the corresponding baseline class.

### Silhouette width

The silhouette width of an observation, in our case a publication, quantifies how well the observation has been clustered (Rousseeuw, 1987). This is done by contrasting coherence to separation: within-cluster dissimilarity is compared to between-cluster dissimilarity. Let $i$ and $j$ be publications such that $i$ and $j$ belong to clusters in a given clustering solution. We define the *dissimilarity between i and j* as 1 if $r_{ij} = 0$, and as 0 if $r_{ij} = 1$ (see Section 4.1.1 for the definition of $r_{ij}$). Informally, the dissimilarity between $i$ and $j$ is 1 if there is no citation relation between *i* and *j*, and the dissimilarity between *i* and *j* is 0 if there is such a relation between *i* and *j*.

For a publication *i* and a clustering solution containing *i*, let *A* be the cluster to which *i* has been assigned, and let *d(i,C)* be the average dissimilarity of *i* to all publications of *C*, where *C* is a cluster of the solution and *C* ≠ *A*. The *silhouette width* for *i*, *s(i)*, is then defined as:

$$s(i) = \frac{b(i) - a(i)}{\max\{a(i), b(i)\}} \tag{7}$$

where *a(i)* is the average dissimilarity of *i* to all other publications of *A*, and $b(i) = \min d(i, C)$. The silhouette width takes values on the interval [-1, 1].

As a clustering solution quality measure, we used the average *s(i)* across all publications in a dataset.

*Probably inaccurate assignments*

We designed a measure, *probably inaccurate assignments* (PIA), with the intention to quantify the extent of the problem of the fractional approach indicated in Section 2. Recall that the problem concerns nodes with many citation relations outside their clusters and only a few relations inside their clusters. PIA is defined, with respect to a given clustering solution, as the number of publications *i* in the solution that satisfy the following three conditions:

a) *i* has at least 20 citation relations,
b) *i* has less than 10% of its citation relations within its cluster, and
c) *i* has a negative silhouette width.

If the three conditions are satisfied, one may state that (i) *i* has sufficiently many citation relations to classified in a proper way, (ii) only a small proportion of the citation relations of *i* are within the cluster of *i*, and (iii) there is at least one other and more proper cluster to which *i* could have been assigned (cf. the definition silhouette width above). Notice that the node in red color in the example of Section 2 stands for a publication, which satisfies the three PIA conditions, given that we assume that its silhouette width is negative.

*4.3.2. Granularity and granularity-quality plots*

We define the *granularity* of a clustering solution as the number of publications divided by the sum of the squared cluster sizes (Waltman et al., 2020). Ideally, and for fairness reasons, clustering solutions compared with regard to the three evaluation measures should have exactly the same granularity. For the five approaches where normalization of direct citations is used, the granularity requirement can be assumed to be approximately satisfied by solutions obtained using different approaches but associated with the same value of the resolution parameter γ. However, the granularity of the clusters obtained from the unnormalized approach deviates somewhat from the granularity of the other approaches. This should be taken into account when the results are interpreted.

We visualize the results by using granularity-quality plots, inspired by earlier, related studies in which granularity-accuracy (GA) plots have been used (Ahlgren et al., 2020; Boyack & Klavans, 2020; Waltman et al., 2020). The use of quality-granularity plots is a way to counteract the difficulty that the granularity requirement referred to in the preceding paragraph is only approximately satisfied. In a granularity-quality plot, the horizontal axis represents granularity (as defined above), whereas the vertical axis represents *M*, where *M* is one of the three evaluation measures used in this study. For a given normalization approach, like fractional, a point in the plot represents the *M* value and granularity of a clustering solution, obtained using a certain resolution value of γ. Further, a line approach (approximate) each point of the normalization approach, where *M* values for granularity values between points are estimated. In this way, the performance of the approaches can be compared at a given granularity level. In our case, the lines are xsplines, i.e. curves drawn relative to control points (Blanc & Schlick, 1995).

## 5 Results

In this section we first present plots showing the skewness of the cluster size distributions resulting from the six normalization approaches in each of the four data sets (Section 5.1). We then present granularity-quality plots for the three evaluation measures (Sections 5.2-5.3).

### 5.1. Skewness

The unnormalized approach results in cluster size distributions that are much more skewed than when normalization is performed (Figure 3). The fractional approach results in the least skewed distribution in most data sets. However, when the granularity increases, the differences between the normalization approaches are very small.

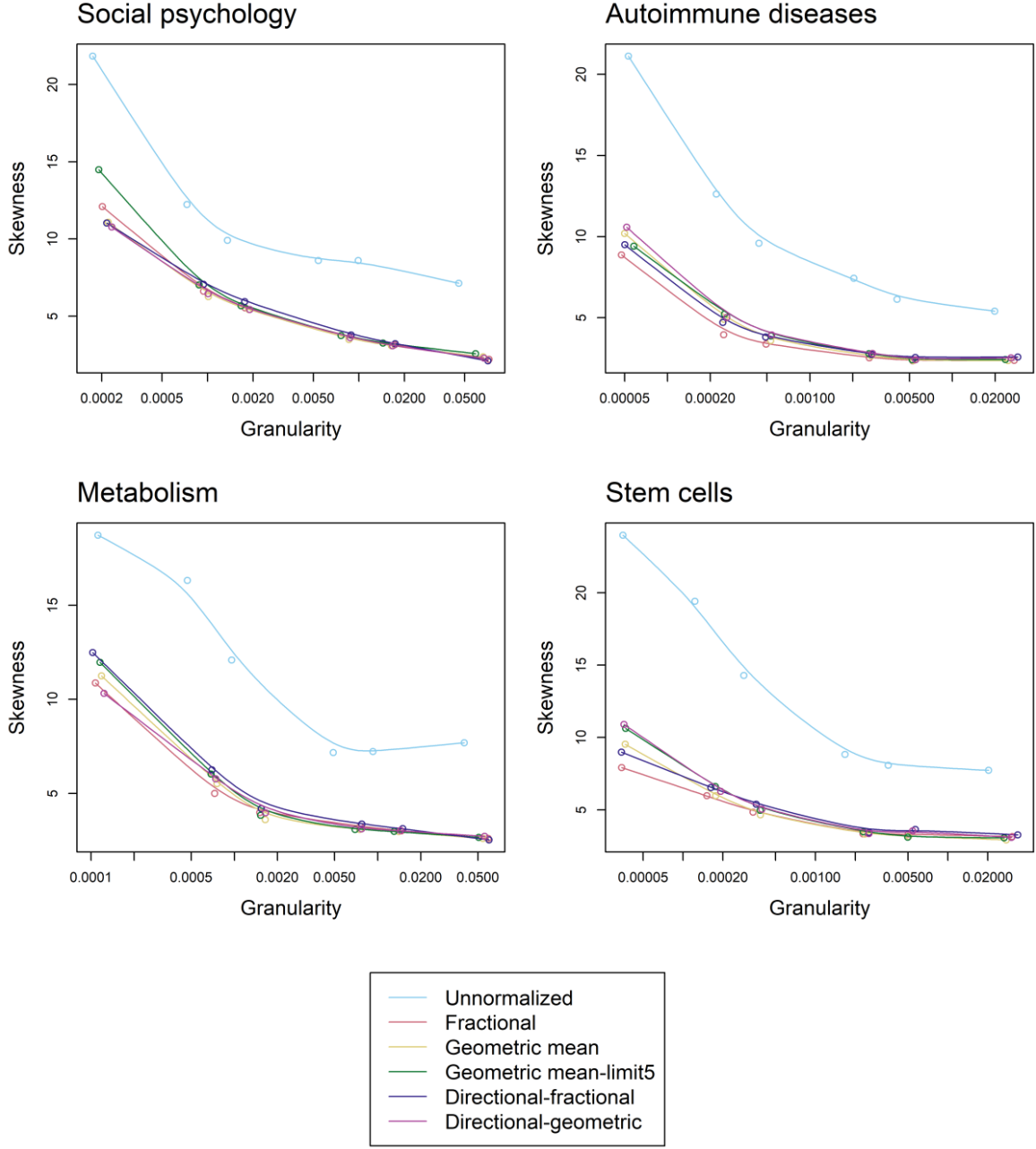

*Figure 3: Skewness of the cluster size distribution (y-axis) by granularity level (x-axis) for the six normalization approaches in the four data sets.*

## 5.2. ARI

The directional-fractional approach resulted in the highest ARI values (Figure 4), but the values does not differ much from the rest of the normalization approaches. If no normalization is performed, the ARI value is notably lower at most granularity levels. The unnormalized approach has higher ARI values for the most granular clustering solutions but its maximal ARI value is lower compared to the maximal ARI values of the other approaches. It should be noted that the unnormalized approach contains high numbers of very small clusters in the most granular clustering solutions. For example, in the most granular clustering solution in "Stem cells", about 13 thousand out of 17 thousand clusters have less than 10 publications. Such feature is unwanted in many practical applications, e.g. if clusters are used for normalization of citation counts.

The ARI value reaches its highest point at a granularity level of about 0.005 in three of the four fields, and even lower for "Social Psychology". Thus, the higher granularity levels capture clusters that have a narrower scope than what is covered in review papers. Such small clusters are probably not very useful in practical applications.

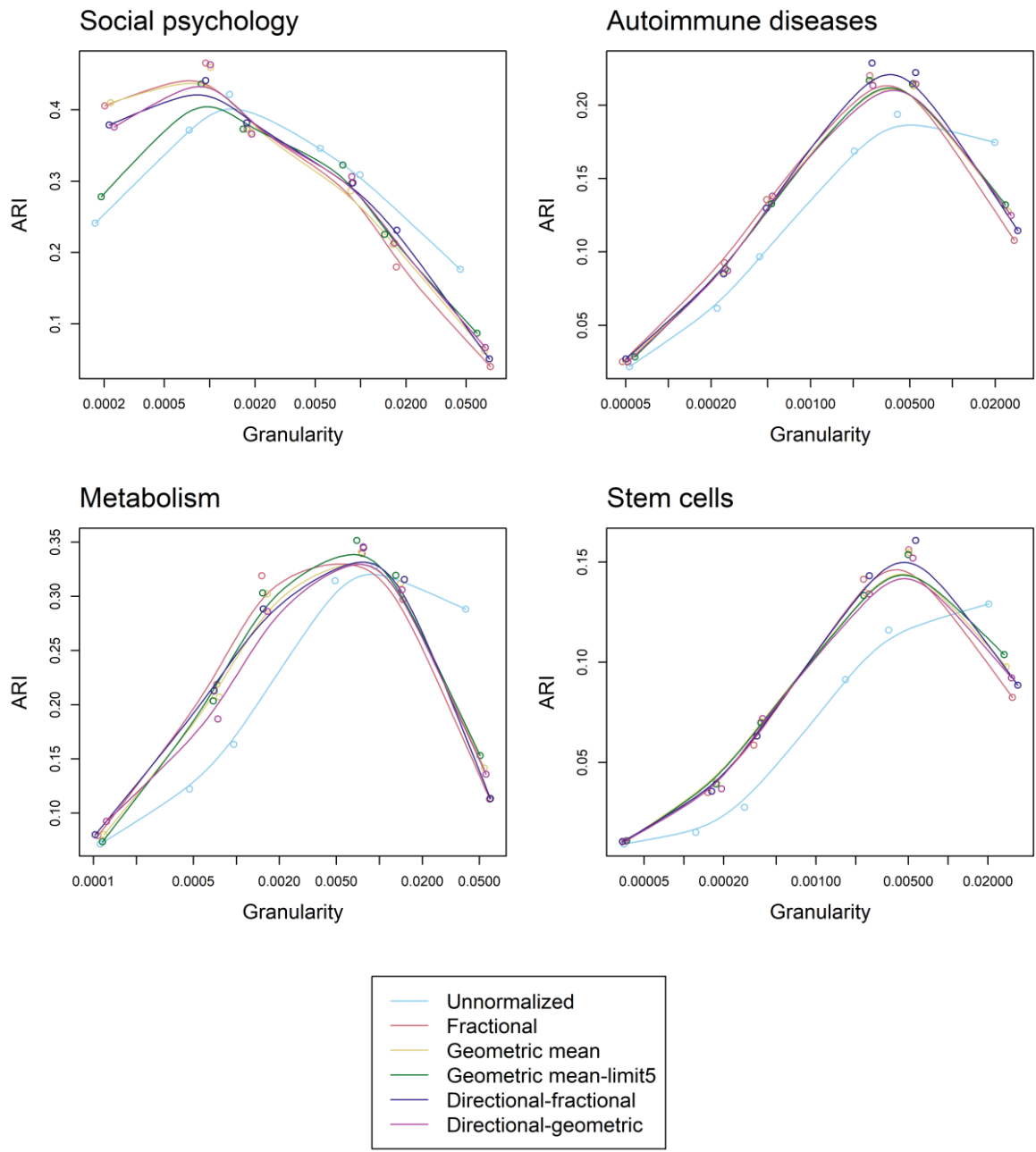

*Figure 4: ARI-values (y-axis) by granularity level (x-axis) for the six normalization approaches in the four data sets.*

5.3. Silhouette width

Also in the case of silhouette width (Figure 5), the values are pretty similar for all of the normalization approaches. The fractional approach has the highest value in most fields and at most granularity levels. The unnormalized approach has lower values at most granularity levels in most fields. The values of the normalization approaches are generally rather close to 1, which suggests that many publications have rather strong connections to another cluster than the one they have been assigned to. This may be an indication of the overlapping nature of research fields.

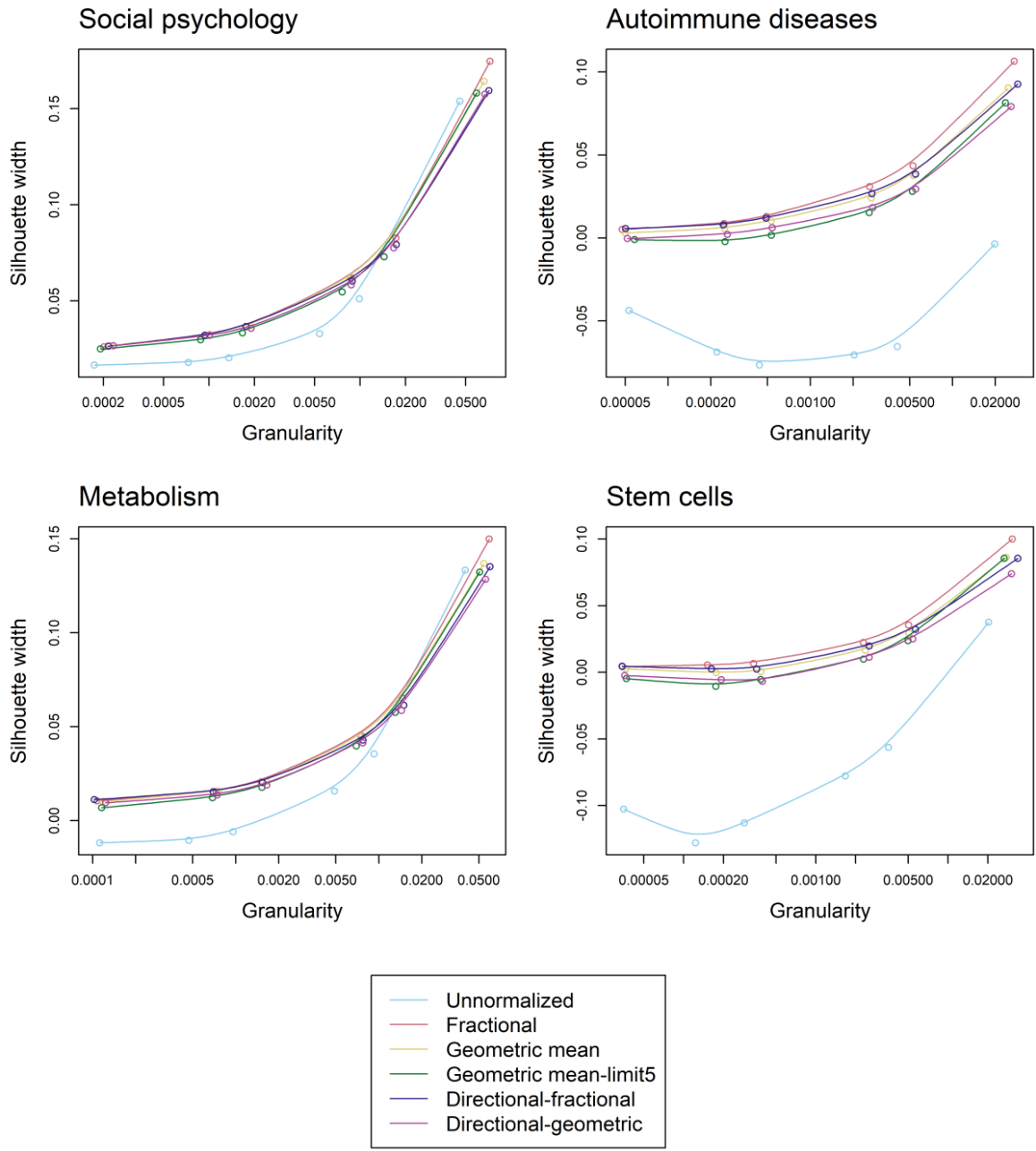

*Figure 5: Silhouette width (y-axis) by granularity level (x-axis) for the six normalization approaches in the four data sets.*

## 5.4. PIA

The unnormalized approach results in the lowest PIA values, i.e., the lowest numbers of inaccurate assignments (Figure 6). The PIA value increases with higher granularity. The two fractional approaches have the highest PIA values, while the geometric and geometric-lim5 approaches have substantially lower values. The results indicate that the geometric approach reduces the problem with inaccurate assignments related to the fractional approach, a problem caused by the high normalized relatedness of a publication, with few relations, with another publication, which is cited by or cites the publication.

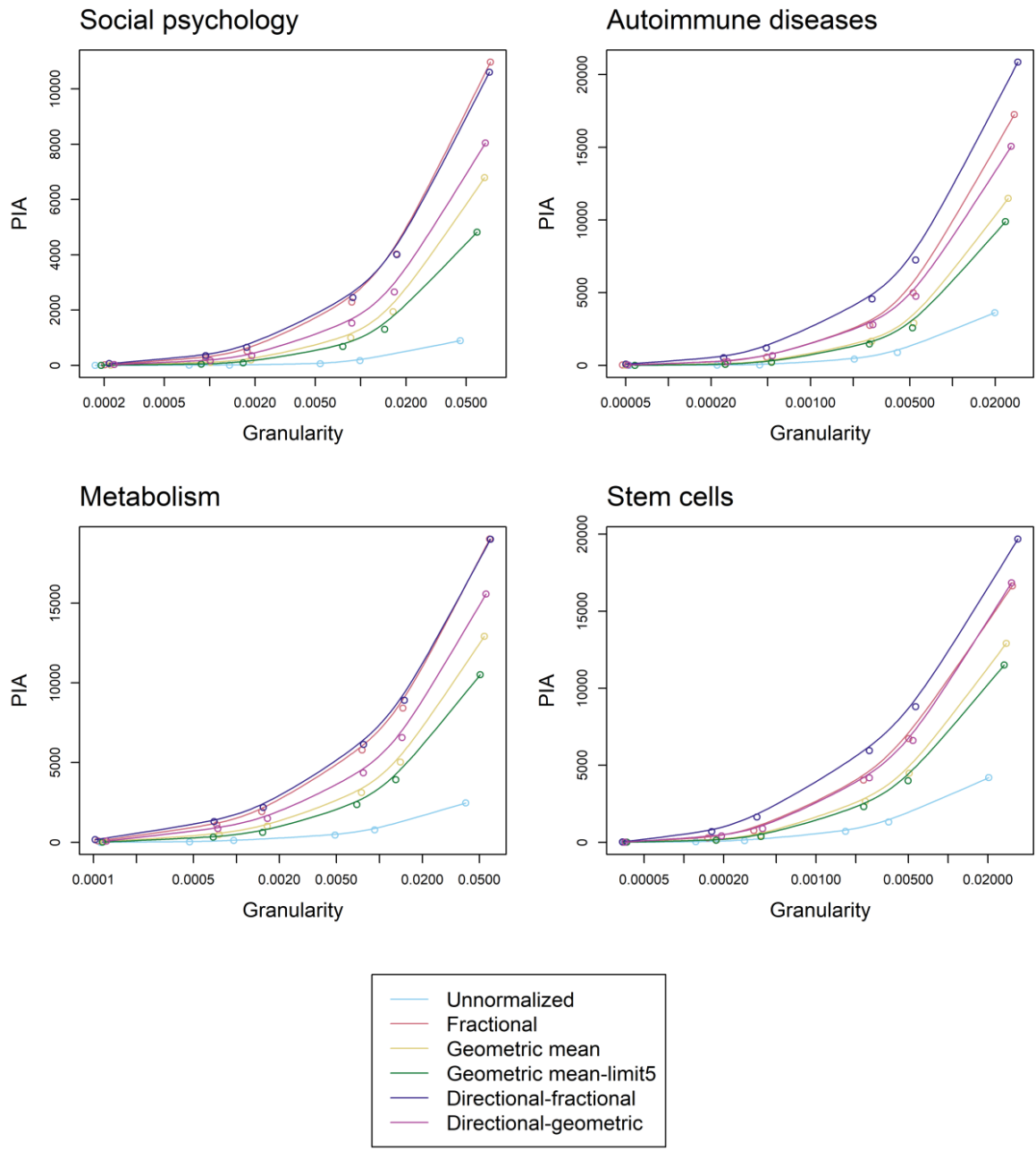

*Figure 6: PIA (y-axis) by granularity level (x-axis) for the six normalization approaches in the four data sets.*

## 6 Discussion and conclusions

The performance of the different normalization approaches is rather similar regarding ARI and silhouette width. The fractional approach, which can be considered as the standard approach for normalization of direct citation relations, performs as well as the other approaches regarding these evaluation measures. The fractional approach also results in cluster size distributions that are among the least skewed. However, the fractional approach has been shown to result in high PIA values. Recall that the PIA measure captures publications assigned to clusters to which they have few relations, despite the facts that they have enough relations to be properly assigned and there exist other clusters to which they have relatively more relations. The geometric approach and geometric-lim5 approach have lower PIA values (especially at higher granularity levels), compared to the approaches fractional, directional-fractional and directional-geometric. The former two approaches may be used to reduce the problem of inaccurate assignment of publications with a modest number of citation relations.

We do not believe that changing from the fractional normalization approach will result in a clustering solution free from poorly connected clusters. Poorly connected clusters may also be a consequence of the clustering algorithm. Park et al. (2023) show in a recent preprint that poorly connected clusters are produced by several of the commonly used community detection algorithms, including the Leiden algorithm for maximizing the Constant Potts Model. They propose a method to remediate poorly connected clusters to improve the connectedness of the clusters in a clustering solution. Such an approach may be combined with a geometric approach for normalization to further reduce the problems of poorly connected clusters and inaccurate assignments. Furthermore, our results support reassignment of publications belonging to small clusters (Waltman & van Eck, 2012). The PIA value increases with higher granularity, indicating that the problem of inaccurate assignments grows with smaller clusters. Reassigning publications in small clusters, clusters with fewer publications than a threshold value, is likely to reduce the problem of poorly connected clusters.

Nonetheless, to accurately assign publications with few citation relations (or even no citation relations), it is necessary to make use of more information. Publications with few citation relations are inevitable difficult to assign to an appropriate cluster. Combining the direct citation approach with a textual-based approach may increase the density in sparse areas of the network. However, such combined approaches have not shown to perform substantially better than a standalone use of direct citations, or extended direct citations, in a couple of previous studies (Ahlgren et al., 2020; Boyack & Klavans, 2020). Furthermore, combined approaches may make interpretation of the clustering solution more difficult in that it becomes less obvious what clusters represent. A direct citation relation implies that the citing authors are aware of the cited publication and explicitly mention this publication in their texts. On the other hand, a textual similarity of two publications occurs when two publications use the same terms in, for example, titles and abstracts, which may happen without awareness of each other's work. This exemplifies the different natures of citation-based and textual similarity. Combining the approaches in one single network may therefore make it unclear how publications in a cluster are related to each other.

Future work may focus on how to address the problem with sparse areas of citation networks. An alternative would be to initially disregard publications with few relations and create a clustering solution including publications with a substantial amount of citation relations. This would create a clustering solution in which clusters represent dense areas of formal communication represented by citations (Sjögårde, 2023). Publications with few citation relations could then be assigned to clusters based on a textual based approach. Future work may also address the performance of clustering approaches that provide overlapping clustering solutions. Such approaches may perform differently in terms of the evaluation measures used in the present work.

In this study, we have compared six approaches to normalization of direct citations with respect to clustering solution quality in four data sets. We conclude that the geometric approach has a similar performance as the fractional approach regarding ARI and silhouette width. However, the results indicate that the geometric approach reduces the problem of inaccurate assignments, and therefore we believe that the geometric approach may be preferred over the fractional approach.

## Data and code availability

Data analyzed in this study is openly available in Zenodo at https://zenodo.org/record/8343758.

Code used for data analysis in this study is openly available in GitHub at https://github.com/petersjogarde/papers/tree/main/normalization_dc_evaluation.

## Author contributions

Peter Sjögårde: Conceptualization; methodology; software; formal analysis; writing—original draft; writing—review & editing; visualization. Per Ahlgren: Conceptualization; methodology; formal analysis; writing—original draft; writing—review & editing.

## Competing interests

The authors declare no competing interests.

## Funding information

Peter Sjögårde was funded by The Foundation for Promotion and Development of Research at Karolinska Institutet.